\documentclass[12pt]{amsart}

\usepackage{graphicx}

\numberwithin{equation}{subsection}
\newfont{\bbd}{msbm10 scaled\magstep1}

\def\l{\lambda}

\let\tilde=\widetilde

\def\id{\hbox{{1}\kern-.25em\hbox{\rm l}}}
\def\one#1{#1^{\raise5pt\hbox{$\scriptstyle\!\!\!\!1$}}\,{}}
\def\two#1{#1^{\raise5pt\hbox{$\scriptstyle\!\!\!\!2$}}\,{}}

\def\bt{B\"acklund transformation}
\def\comment#1{}

\def\Ref#1{(\ref{#1})}
\def\?{(?)\marginpar{|?}}

\def\beq{\begin{equation}}
\def\eeq{\end{equation}}
\def\be{\begin{displaymath}}
\def\ee{\end{displaymath}}
\def\bea{\begin{eqnarray}}
\def\eea{\end{eqnarray}}
\def\bmat{\left(\begin{array}}
\def\emat{\end{array}\right)}

\newcounter{subequation}[equation]
\makeatletter
\expandafter\let\expandafter
\reset@font\csname reset@font\endcsname
\def\subeqnarray{\arraycolsep1pt
    \def\@eqnnum\stepcounter##1{\stepcounter{subequation}%
        {\reset@font\rm(\theequation\alph{subequation})}}
\jot5mm     \eqnarray}

\makeatother

\begin{document}
\title[Separation of variables and B\"acklund transformations]%
  {Separation of variables and B\"acklund transformations for the
symmetric Lagrange top}
\author{Vadim B. Kuznetsov}
  \address{Department of Applied Mathematics,
          University of Leeds,
          Leeds LS2 9JT, UK}
  \email{V.B.Kuznetsov@leeds.ac.uk}
\author{Matteo Petrera}
  \address{Dipartimento di Fisica 'E Amaldi', Universit\'a degli Studi 'Roma Tre',
and Istituto Nazionale di Fisica Nucleare, Sezione di Roma Tre, 
Via della Vasca Navale 84, Rome, Italy}
  \email{petrera@fis.uniroma3.it}
\author{Orlando Ragnisco}
\address{Dipartimento di Fisica 'E Amaldi', Universit\'a degli Studi 'Roma Tre', 
and Istituto Nazionale di Fisica Nucleare, Sezione di Roma Tre, 
Via della Vasca Navale 84, Rome, Italy}
 \email{ragnisco@fis.uniroma3.it}
\begin{abstract}
We construct the 1- and 2-point integrable maps (B\"acklund transformations)
for the symmetric Lagrange top. We show that the Lagrange top has the same algebraic Poisson
structure that belongs to the $sl(2)$ Gaudin magnet.
The 2-point map leads to a real time-discretization
of the continuous flow. Therefore, it provides an integrable numerical scheme
for integrating the physical flow. We illustrate the construction by few pictures
of the discrete flow calculated in MATLAB.
\end{abstract}

\maketitle
\tableofcontents
%%%%%%%%%%%%%%%%%%%%%%%%%%%%%%%%%%%%%%%%%%%%%%%%%%
\section{Introduction: the symmetric Lagrange top} 
\setcounter{equation}{0}
%%%%%%%%%%%%%%%%%%%%%%%%%%%%%%%%%%%%%%%%%%%%%%%%%%
\noindent
The Lagrange top is an integrable case of rotation of a rigid 
body around a fixed point
in a homogeneous gravitational field, characterized by the following 
conditions \cite{Au,Gav2}: the
rigid body
is rotationally symmetric, i.e. two of its three principal moments of inertia
coincide,
and the fixed point lies on the axis of the rotational symmetry. A standard
form of the corresponding equations of motion is given by 
the Euler-Poisson equations:
\begin{equation}
\left\{ \begin{array}{ll}
\dot{J}=P \times x,\\
\dot{x}=J \times x.
\end{array} \right.
\label{eqqi}
\end{equation}
Here $J=(J_1,J_2,J_3) \in \mathbb{R}^3$ is the vector of angular 
momentum of the
body,
$P=(0,0,\alpha) \in \mathbb{R}^3$ is the constant vector 
along the gravity field and
$x=(x_1,x_2,x_3) \in \mathbb{R}^3$ is the vector pointing 
from the fixed point to the
center of mass.

The symmetric Lagrange top is an integrable system 
with 2 degrees of freedom and the Hamiltonian
\begin{equation}
\mathcal{H}=\frac{1}{2}(J_1^2+J_2^2+J_3^2)+\alpha x_3,
\label{H}\end{equation}
where  $J_k,x_k$, $k=1,2,3$ are six generators
of the Lie-Poisson $e(3)$ algebra defined by the
following Poisson brackets: 
\begin{equation}
\{J_k, J_l\}=J_m, \qquad \{J_k,x_l\}=x_m,
\qquad
\{x_k,x_l\}=0,
\label{2}\end{equation}
 $(klm)$ is a cyclic permutation of $(123)$. 

We will also use the complex conjugated variables $J_\pm=J_1\pm i J_2$
and $x_\pm=x_1\pm i x_2$, which have the brackets
\begin{equation}
\{J_{3},J_{\pm}\}=\mp iJ_{\pm}, \qquad
\{J_{+},J_{-}\}=-2iJ_{3}, \qquad
\{J_{3},x_{\pm}\}=\{x_{3},J_{\pm}\}=\mp ix_{\pm},
\label{3-5}\end{equation}
\be
\{J_{+},x_{-}\}=\{x_{+}, J_{-}\}=-2ix_{3}, \qquad
\{J_{3},x_{3}\}=\{J_{+},x_{+}\}=\{J_{-},x_{-}\}=0,
\ee
\be
\{x_{k},x_{l}\}=0,  \qquad k,l=\pm,3.
\ee

The Lie-Poisson bracket \Ref{2} have two Casimir functions
\begin{equation}
C_1 \equiv  \sum_{k=1}^{3} x_k^2,\qquad C_2 \equiv
\sum_{k=1}^{3} x_k J_k .
\end{equation}
Fixing their values one gets a generic symplectic leaf
\begin{equation}
 \mathcal{O}_{c_1,c_2} \equiv \{x,J| \;C_1 = c_1, \;C_2=c_2   \},
\end{equation}
which is a four-dimensional symplectic manifold. Hereafter
we take $c_1=1$ and $c_2 = \ell$, corresponding to a unit vector $x$
and a fixed projection $\ell$ of the angular momentum $J$ on the 
vector $x$.

Two commuting integrals of motion (Hamiltonians) 
of the symmetric Lagrange top
are respectively,
\begin{equation}
\mathcal{H}=\frac{1}{2}(J_1^2+J_2^2+J_3^2)+\alpha x_3 
\qquad \mbox{and}\qquad J_3,
\qquad \{\mathcal{H},J_3\}=0.
\end{equation}
Obviously, the conservation of $J_3$ is a direct consequence 
of the invariance under
rotation about the
direction of the gravity field.
Using complex generators of $e(3)$ we can write the Hamiltonian as
\begin{equation}
\mathcal{H}=\frac{1}{2}(J_+ J_-+J_3^2)+\alpha x_3 .
\end{equation}
%%%%%%%%%%%%%%%%%%%%%%%%%%%%%%%%%%%%%%%%%%%%%%
\section{From the $sl(2)$ Gaudin magnet  to the symmetric Lagrange top} 
\setcounter{equation}{0}%%
%%%%%%%%%%%%%%%%%%%%%%%%%%%%%%%%%%%%%%%%%%%%%%
\noindent
The symmetric Lagrange top can be derived from 
the $sl(2)$ Gaudin magnet \cite{G1,G2}, which has  the $2\times 2$ Lax matrix
\begin{equation}
L_{\mathcal{G}}(u)=\sum_{j=1}^{n}\frac{1}{u-a_{j}}
\left(\begin{array}{cc}
s_{j}^{3} & s_{j}^{-} \\
s_{j}^{+} & -s_{j}^{3}
\end{array}\right)+
\alpha
\left(\begin{array}{cc}
1 & 0 \\
0 & -1
\end{array}\right)
=\begin{pmatrix} A_{\mathcal{G}}(u)&B_{\mathcal{G}}(u)\cr
C_{\mathcal{G}}(u)&-A_{\mathcal{G}}(u)  \end{pmatrix},
\label{10}\end{equation}
\begin{equation}
A_{\mathcal{G}}(u)=\alpha+\sum_{j=1}^{n}\frac{s_{j}^{3}}{u-a_{j}}\,, \qquad
B_{\mathcal{G}}(u)=\sum_{j=1}^{n}\frac{s_{j}^{-}}{u-a_{j}}\,, \qquad
C_{\mathcal{G}}(u)=\sum_{j=1}^{n}\frac{s_{j}^{+}}{u-a_{j}}\,,
\label{rat-fun}
\end{equation}
where $a_j \in \mathbb{C}$ and $\alpha\in \mathbb{R}$ are the parameters of the model
and $u \in \mathbb{C}$ is the spectral parameter. The parameter 
$\alpha $ has the meaning of magnetic field's intensity.
Local variables $s_{j}^{3},s_{j}^{\pm}$, $j=1,...,n$, are the generators 
of the direct sum of $n$ $sl(2)$ spins with the
following Poisson brackets:
\begin{equation}
\lbrace{s_{j}^{3},s_{k}^{\pm} \rbrace}=\mp i \delta_{jk}s_{k}^{\pm}, \qquad
\lbrace{s_{j}^{+},s_{k}^{-} \rbrace}=-2i \delta_{jk}s_{k}^{3}.
\end{equation}
We denote the Casimir operators (spins) as $s_{j}$:
\begin{equation}
s_{j}^{2}=(s_{j}^{3})^{2}+s_{j}^{+}s_{j}^{-}.
\end{equation}
Fixing $s_{j}$'s we go to a symplectic leaf where the Poisson bracket
is non-degenerate, so that the symplectic manifold is a collection of $n$
spheres.

The Lax matrix \Ref{10} satisfies the linear $r$-matrix Poisson algebra:
\begin{equation}
\lbrace{\stackrel{1}{L}_{\mathcal{G}}(u),\stackrel{2}{L_{\mathcal{G}}}(v) \rbrace}=
[r(u-v),\stackrel{1}{L}_{\mathcal{G}}(u)+\stackrel{2}{L}_{\mathcal{G}}(v)], \quad
\stackrel{1}{L}_{\mathcal{G}}=L_{\mathcal{G}} \otimes 
\left(\begin{matrix}
1&0\cr 0&1
\end{matrix}\right), \quad
\stackrel{2}{L}_{\mathcal{G}}=\left(\begin{matrix}
1&0\cr 0&1
\end{matrix}\right)
 \otimes L_{\mathcal{G}},
\label{14}\end{equation}
with the permutation matrix as the $r$ matrix:
\begin{equation}
r(u-v)=\frac{i}{u-v}\left(\begin{array}{cccc}
1 & 0 & 0 & 0\\
0 & 0 & 1 & 0\\
0 & 1 & 0 & 0\\
0 & 0 & 0 & 1
\end{array}\right).
\label{15}\end{equation}
Equation \Ref{14} is equivalent to the following Poisson brackets for the rational
functions
$A_{\mathcal{G}}(u)$, $B_{\mathcal{G}}(u)$, $C_{\mathcal{G}}(u)$
\Ref{rat-fun}:
\begin{eqnarray}
\{A_{\mathcal{G}}(u),A_{\mathcal{G}}(v)\}&=&\{B_{\mathcal{G}}(u),B_{\mathcal{G}}(v)\}
=
\{C_{\mathcal{G}}(u),C_{\mathcal{G}}(v)\}=0,\\
\{A_{\mathcal{G}}(u),B_{\mathcal{G}}(v)\}&=&\frac{i}{u-v}
[B_{\mathcal{G}}(v)-B_{\mathcal{G}}(u)],\\
\{A_{\mathcal{G}}(u),C_{\mathcal{G}}(v)\}&=&\frac{i}{u-v}
[C_{\mathcal{G}}(u)-C_{\mathcal{G}}(v)],\\
\{C_{\mathcal{G}}(u),B_{\mathcal{G}}(v)\}&=&\frac{2i}{u-v}
[A_{\mathcal{G}}(u)-A_{\mathcal{G}}(v)].
\end{eqnarray}

The non-linear dynamics defined by the equations \Ref{eqqi} is linearised
on the Jacobian of the hyperelliptic spectral curve $\Gamma_{\mathcal{G}}$ of  genus $n-1$,
\begin{equation}
\Gamma_{\mathcal{G}}: \qquad \det(L_{\mathcal{G}}(u)- v)=0,
\end{equation}
which can be brought into the form
\begin{equation}
v^{2}=A_{\mathcal{G}}^{2}(u)+B_{\mathcal{G}}(u)C_{\mathcal{G}}(u)=\alpha^{2}+
\sum_{j=1}^{n} \left(\frac{H_{j}}{u-a_{j}}+
\frac{s_{j}^{2}}{(u-a_{j})^{2}} \right).
\end{equation}
The Hamiltonians $H_j$ above are given by
\begin{equation}
H_{j}=\sum_{k \neq j} \frac{2s_{j}^{3}s_{k}^{3}+s_{j}^{+}s_{k}^{-}+
s_{j}^{-}s_{k}^{+}} {a_{j}-a_{k}}
+2\alpha s_{j}^{3}.
\end{equation}
These are integrals of motion of the $sl(2)$ Gaudin magnet, which are
Poisson commuting:
\begin{equation}
\lbrace{H_{j},H_{k} \rbrace}=0 \qquad j,k=1,...,n.
\end{equation}

Take a 2-site Gaudin spin chain for which $n=2$ and the phase space is
the direct sum of two $sl(2)$ spins.
We use the notations $s_j^1$ and $s_j^2$ defined by
$s_{j}^{\pm}=s_j^1 \pm i s_j^2$. Introduce a new matrix, called $L(u)$, 
as follows:
\begin{equation}
L(u) \equiv  i L_{\mathcal{G}}^{(n=2)}(u)=
\frac{1}{u-a_{1}}\, L_1 +
\frac{1}{u-a_{2}}\,L_2
+ \alpha \left(\begin{array}{cc}
i & 0 \\
0 & -i
\end{array}\right),
\label{24}\end{equation}
where
\begin{equation}
L_k \equiv \left(\begin{array}{cc}
 is_{k}^{3} & s_{k}^{2}+is_{k}^{1}  \\
-s_{k}^{2}+is_{k}^{1} & -is_{k}^{3}
\end{array} \right) \in su(2), \qquad k=1,2.
\label{25}\end{equation}
Let us consider the following Lie algebras' isomorphism:
\begin{equation}
su(2) \oplus su(2) \cong o(4).
\end{equation}
The In\"on\"u-Wigner contraction from the rotation group $O(m+1)$ to the
Euclidean group $E(m)$ \cite{IW}, for $m=3$, allows us to obtain the Lie-Poisson
algebra $e(3)$ for the classical Lagrange top, with six generators $J_k,x_k$,
$k=\pm,3$.

Let us introduce the contraction parameter $\epsilon \in (0,1]$.
Impose in the equation \Ref{24} the coalescence
$a_{2} =a_{1}+\epsilon$, $\epsilon \rightarrow 0$ and redefine
$u-a_{1} \equiv u$. Then the Lax matrix reads
\begin{equation}
L(u)=\frac{1}{u}\,(L_{1}+L_{2})+\frac{1}{u^{2}}\,\epsilon L_{2}+
\alpha
\left(\begin{array}{cc}
i & 0 \\
0 & -i
\end{array}\right)
+O(\epsilon^{2}).
\end{equation}
In order to have a $2\times 2$ Lax matrix for the Lagrange top we have to
control that $L_{1}+L_{2}$ and $\epsilon L_{2}$ play the role of the
following matrices respectively:
\begin{equation}
J \equiv i \left(\begin{array}{cc}
J_{3} & J_{-}  \\
J_{+} & -J_{3}
\end{array} \right), \qquad
x \equiv i \left(\begin{array}{cc}
x_{3} & x_{-}  \\
x_{+} & -x_{3}
\end{array} \right),
\end{equation}
or, in other words, we have to check the following Poisson morphisms:
\begin{equation}
s_{1}^{3}+s_{2}^{3} \cong J_{3}, \qquad
s_{1}^{+}+s_{2}^{+} \cong J_{+}, \qquad
s_{1}^{-}+s_{2}^{-} \cong J_{-},
\label{29}\end{equation}
\be
\epsilon s_{2}^{3} \cong x_{3}, \qquad
\epsilon s_{2}^{+} \cong x_{+}, \qquad
\epsilon s_{2}^{-} \cong x_{-}.
\ee
With a direct calculation of Poisson brackets it is possible to prove 
that morphisms \Ref{29} actually hold.

Therefore we obtain
the $2\times 2$ Lax matrix for the Lagrange top
\begin{equation}
L(u)=\begin{pmatrix} A(u)&B(u)\cr C(u)&-A(u)  \end{pmatrix}=i
\left(\begin{matrix} \frac{J_3}{u}+\frac{x_3}{u^2}+\alpha
&  \frac{J_-}{u}+\frac{x_-}{u^2}
&\cr  \frac{J_+}{u}+\frac{x_+}{u^2}
& -\frac{J_3}{u}-\frac{x_3}{u^2}-\alpha
 \end{matrix}\right),
\label{31}\end{equation}
We want to remark that in the Lagrange case the real parameter $\alpha$
describes the gravitational field's intensity.

The spectral curve of the Lagrange top is the elliptic curve
$\Gamma:\;\det(L(u)-v)=0$,
\begin{equation}
\Gamma:\qquad -v^2=\frac{1}{u^4}+\frac{2\ell}{u^3}+\frac{2\mathcal{H}}{u^2}+
\frac{2J_3}{u}+\alpha^2.
\label{Ga}
\end{equation}

The Lie algebra contraction $o(4) \rightarrow e(3)$ does not change the
$r$-matrix structure of the model, that remains in the form \Ref{14}--\Ref{15}
with the Lax matrix given by \Ref{31}.
%%%%%%%%%%%%%%%%%%%%%%%%%%%%%%%%%%%%%%%%%%%%%%
\section{Separation of variables} 
\setcounter{equation}{0}%%
%%%%%%%%%%%%%%%%%%%%%%%%%%%%%%%%%%%%%%%%%%%%%%
\noindent
In this Section we construct the simplest separation of variables
for the symmetric Lagrange top with the Lax matrix \Ref{31}.
The details of the approach can be found in \cite{Skl38, K, KNS}.

The basic separation has only one pair $(u_1,v_1)$ of separation variables
belonging to  the spectral curve $\Gamma$ \Ref{Ga}. It corresponds to the standard normalisation vector
$\alpha_0=(1,0)$ and it is defined by the equations
\begin{equation}
(1,0) (L(u_1)-v_1)^\wedge=0,
\label{34}\end{equation}
where $(\cdot)^\wedge$ denotes the adjoint matrix. Equation \Ref{34} is the equation for
the pole $(u_1,v_1)$
of the Baker-Akhiezer function $\Psi$, which is defined as a properly normalised eigenfunction
of the Lax matrix:
\begin{equation}
L(u) \Psi = v \Psi, \qquad (u,v) \in \Gamma,
\end{equation}
\begin{equation}
\alpha_0 \cdot \Psi =1.
\end{equation}
It is easy to see that the
equation \Ref{34} gives the following separation variables:
\begin{equation}
B(u_1)=0, \qquad v_1=-A(u_1).
\end{equation}
Explicitly, the pair of the canonical separation variables is
\begin{equation}
u_1=-\frac{x_-}{J_-}\,, \qquad
v_1=i\, \frac{J_-}{x_-}\left(J_3- \frac{J_-}{x_-}x_3\right).
\end{equation}
In order to define the map between the initial and separation
variables, one has to add an extra pair of canonical variables. This pair of variables is taken from the
asymptotics ($u \rightarrow \infty$) of the elements $A(u)$ and $B(u)$ of the Lax matrix. We obtain
\begin{equation}
u_2=J_-, \qquad
v_2=-i\,\frac{J_3}{J_-}.
\end{equation}
Indeed, using the Lie-Poisson algebra $e(3)$ \Ref{3-5},
it is easy to check  that the new variables are canonical:
\begin{equation}
\{u_k, u_l\}=\{v_k, v_l\}=0, \qquad \{v_k,u_l\}=\delta_{kl}, \qquad k,l=1,2.
\label{bracket2}\end{equation}
Remark that the new separation variables $(u_1,u_2,v_1,v_2)$ are defined in a complex
domain.
We can write down the complex generators of $e(3)$ in terms of the separation
variables:
\begin{eqnarray}
J_{-}&=&u_2,\\
J_{+}&=&-\frac{2\ell}{u_1u_2}+v_2^2u_2-\frac{1}{u_1^2u_2}-
\frac{u_1^2v_1^2}{u_2},\nonumber\\
J_{3}&=&iv_2u_2,\nonumber\\
x_{-}&=&-u_1u_2,\nonumber\\
x_{+}&=&-\frac{1}{u_1u_2}-v_2^2u_1u_2+2v_1v_2u_1^2- \frac{u_1^3v_1^2}{u_2},\nonumber\\
x_{3}&=&i u_1(u_1 v_1-u_2v_2).
\nonumber\end{eqnarray}
The separation equations are:
\begin{eqnarray}
\mathcal{H} &=& u_1^2\left(i\alpha v_1-\frac{v_1^2}{2} \right)-\frac{\ell}{u_1}
-\frac{1}{2u_1^2}-\alpha J_3u_1, \\
J_3 &=& iv_2u_2.
\end{eqnarray}

One can use the above separation of variables to integrate the model
in terms of elliptic functions. Although, because the separation variables
are complex, it is not easy to apply the reality conditions. In the next
Section we describe the standard integration of the symmetric Lagrange
top.

We finish this Section by presenting the canonical separating transform 
through a generating function.

Let us fix a special representation of the $e(3)$ algebra
in terms of the Darboux coordinates $q_j,p_j,
j=1,2$:
\begin{eqnarray}
J_{-}&=&q_1,\\
J_{+}&=&q_1p_{1}^{2}+2q_2p_1p_2-2i\ell p_1-2ip_2,\nonumber\\
J_{3}&=&-i(q_1p_1+q_2p_2)- \ell,\nonumber\\
x_{-}&=&q_2,\nonumber\\
x_{+}&=&q_2p_{1}^{2}-2ip_1,\nonumber\\
x_{3}&=&-iq_2p_1-1,
\nonumber\end{eqnarray}
\begin{equation}
\{q_k, q_l\}=\{p_k, p_l\}=0, \qquad \{q_k,p_l\}=\delta_{kl}, \qquad k,l=1,2.
\end{equation}
Notice that in this representation the variables $J_-$ and $x_-$
do not depend on the momenta and the variables $J_3$ and $x_3$ are linear in the
momenta.

Finally, the canonical transformation $(q_1,q_2,p_1,p_2) \rightarrow
(u_1,u_2,v_1,v_2)$ is defined by the generating
function
\begin{equation}
F(q_1,q_2 \mid v_1,v_2) =i\,\frac{q_1}{q_2}+i\ell \ln q_2 -q_1v_2
+\frac{q_2v_1}{q_1}\,,
\end{equation}
that is
\beq
p_i = \frac{\partial F(q_1,q_2 \mid v_1,v_2)}{\partial q_i}\,, \qquad
u_i = -\frac{\partial F(q_1,q_2 \mid v_1,v_2)}{\partial v_i}\,,\qquad i=1,2.
\eeq
%%%%%%%%%%%%%%%%%%%%%%%%%%%%%%%%%%%%%%%%%%%%%%%%%%
\section{Integrating the model} 
\setcounter{equation}{0}%%
%%%%%%%%%%%%%%%%%%%%%%%%%%%%%%%%%%%%%%%%%%%%%%%%%%
\noindent
The standard integration of the Lagrange top is performed through Eulerian variables
which
allow the reduction to one degree of freedom by using the first order
integral $J_3$ to separate out the angle $\varphi$ and its
conjugated momentum $p_\varphi$ \cite{Ar}.

Let us introduce the Eulerian angles $(\theta,\varphi)$, $\theta \in [0,\pi]$,
$\varphi \in [0,2 \pi)$, to parametrize
the unit sphere $x_1^2+x_2^2+x_3^2=1$, which together
with the conjugated momenta $(p_\theta,p_\varphi)$ give
a representation of the $e(3)$ generators in terms of
Darboux coordinates:
\begin{eqnarray}
J_1&=&p_\theta\cos\varphi+\frac{\ell-p_\varphi\cos\theta}{\sin\theta}\,
\sin\varphi,\qquad\quad
x_1=\sin\varphi\sin\theta,\label{59}\\
J_2&=&-p_\theta\sin\varphi+\frac{\ell-p_\varphi\cos\theta}{\sin\theta}\,
\cos\varphi,\qquad\,
x_2=\cos\varphi\sin\theta,\nonumber\\
J_3&=&p_\varphi,\qquad\qquad\qquad\qquad
\qquad\qquad\qquad x_3=\cos\theta,
\nonumber\end{eqnarray}
\begin{equation}
\{p_\theta,\theta\}=\{p_\varphi,\varphi\}=1.
\label{62}\end{equation}
One immediately
arrives at two one-dimensional equations
\begin{eqnarray}
\mathcal{H}&=&\frac{p_\theta^2}{2}+\frac{p_\varphi ^2-2\ell p_\varphi
\cos\theta+\ell^2}{2 \sin^2\theta}+
\alpha \cos \theta,\\
J_3&=&p_\varphi,
\end{eqnarray}
which lead to integration of the model in terms of  the
Weierstrass elliptic function $\wp(x)$. The time evolution of the angles and of their conjugated
momenta is the
following:
\begin{eqnarray}
\theta &=& \arccos
\left[ \frac{E}{3 \alpha} +
\wp \left( \sqrt{ \frac{\alpha}{2}}\;t \right) \right], \\
\varphi &=& \int \frac{m- 2 \ell \left[ \frac{E}{3 \alpha} +
\wp \left( \sqrt{ \frac{\alpha}{2}}\;t \right) \right]}
{1- \left[ \frac{E}{3 \alpha} +
\wp \left( \sqrt{ \frac{\alpha}{2}}\;t \right) \right]^2}dt, \nonumber\\
p_\theta&=& \frac{d}{dt} \left\{ \arccos
\left[ \frac{E}{3 \alpha} +
\wp \left( \sqrt{ \frac{\alpha}{2}}\;t \right) \right]   \right\},\nonumber\\
p_\varphi &=& m.
\nonumber\end{eqnarray}
%%%%%%%%%%%%%%%%%%%%%%%%%%%%%%%%%%%%%%%%%%%%%%
\section{B\"acklund transformations (BT's)} 
\setcounter{equation}{0}%%
%%%%%%%%%%%%%%%%%%%%%%%%%%%%%%%%%%%%%%%%%%%%%%
\noindent
In this paper, following the approach of \cite{KS5,KV00}, we look at the 
B\"acklund transformations (BT's) for finite-dimensional
(Liouville)
integrable systems as special canonical transformations, thereby taking a Hamiltonian
point of
view. Such BT's are defined as symplectic, or
more
generally Poisson, integrable maps which are explicit maps (rather than implicit
multivalued correspondences) and which can be viewed as time discretizations of
particular
continuous flows. 

The most characteristic properties of such maps are: (i) a BT preserves the
same set of integrals of motion as does the continuous flow which it discretizes,
(ii) it 
depends on a B\"acklund parameter $\lambda$ that specifies the corresponding
shift on a Jacobian or on a generalized Jacobian \cite{KV00,Gav1}, and (iii) a spectrality
property holds with respect
to $\lambda$ and to the conjugate variable $\mu$, which means that the point
$(\l,\mu)$ belongs
to the spectral curve \cite{KS5,KV00}.
Explicitness makes these maps purely iterative, while the importance of the parameter
$\l$
is that it allows for an adjustable discrete time step. The spectrality property is
related with the simplecticity of the map \cite{KV00}.

In this paper we construct BT's for the symmetric Lagrange top starting
from the results obtained in \cite{HKR}. As shown in previous sections, the Lagrange
top has the
same algebraic Poisson structure that belongs to the Gaudin magnet. This allows to
choose the same
ansatz for the matrix $M(u)$ that has been used in \cite{HKR}.

We have here to remark that an elegant alternative approach to integrable time discretizations 
of continuous hamiltonian flows has been carried out by Yu. B. Suris and A.I. Bobenko \cite{SB,SU}. 
In particular, in \cite{SB} they construct a discrete time Lagrange top: notice howewer
that their discrete Lax matrix is deformed with respect to the continuous one, which is 
recovered in the limit when the time step goes to zero.

%%%%%%%%%%%%%%%%%%%%%%%%%%%%%%%%%%%%%%%%%%%%%%%%%%%%%
\subsection{One-point BT} 
%%%%%%%%%%%%%%%%%%%%%%%%%%%%%%%%%%%%%%%%%%%%%%%%%%%%%%
\noindent
A (one-point) B\"acklund transformation for the Lagrange top is equivalent to the
following similarity transform on the Lax matrix $L(u)$:
\begin{equation}
L(u)  \longmapsto M(u;\lambda) L(u)M^{-1}(u,\lambda) \qquad \forall u, \qquad \lambda \in \mathbb{C},
\end{equation}
with some non-degenerate $2 \times2$ matrix $M(u,\lambda)$, simply because a BT should
preserve
the spectrum of $L(u)$. The parameter $\lambda \in \mathbb{C}$ is called a B\"acklund parameter of
the transformation.
Let us introduce new $~ \tilde {}$ -notations for the updated variables:
\begin{equation}
\tilde L(u)=\begin{pmatrix} \tilde A(u)&\tilde B(u)\cr \tilde C(u)&-\tilde A(u)
\end{pmatrix}=i
\left(\begin{matrix} \frac{\tilde J_3}{u}+\frac{\tilde x_3}{u^2}+\alpha
&  \frac{\tilde J_-}{u}+\frac{\tilde x_-}{u^2}
&\cr  \frac{\tilde J_+}{u}+\frac{\tilde x_+}{u^2}
& -\frac{\tilde J_3}{u}-\frac{\tilde x_3}{u^2}-\alpha
 \end{matrix}\right).
\label{lax-matrix2}
\end{equation}
We are looking for a Poisson map that intertwines two Lax matrices 
$L(u)$ and $\tilde
L(u)$:
\begin{equation}
M(u;\lambda)L(u)=\tilde L(u)M(u;\lambda) \qquad \forall u.
\label{71}\end{equation}
Let us take
\begin{equation}
M(u;\lambda)=\left(\begin{array}{cc}
u-\lambda +pq& p \\
q & 1
\end{array}\right),\qquad \det M(u;\lambda) =u - \l.
\label{ansatz}
\end{equation}
Let us stress that the number of zeros of $\det M$ is the number of essential B\"acklund
parameters.
Here the variables $p$ and $q$ are indeterminate dynamical variables. 

The ansatz \Ref{ansatz} for the matrix $M$ came from the simplest
$L$-operator of the quadratic $r$-matrix algebra
\begin{equation}
\lbrace{\stackrel{1}{L}(u),\stackrel{2}{L}(v) \rbrace}=
[r(u-v),\stackrel{1}{L}(u)\stackrel{2}{L}(v)],
\label{74}\end{equation}
with the same $r$-matrix \Ref{15}. Notice that we have a peculiar algebraic situation,
as in Gaudin models:
a Lax matrix which satisfies a {\it linear} $r$-matrix algebra requires a $M$ matrix
which comes from a {\it quadratic} $r$-matrix algebra. Recall that in the Toda lattice 
\cite{KS5} and in the DST model \cite{KSS98} both
$L$ and $M$ are derived from the same quadratic $r$-matrix algebra. This fact
shall show up in the construction of the quantum analogue of these B\"acklund
transformations.

Comparing the asymptotics in $u\rightarrow\infty$ in both sides of \Ref{71} we readily get
\begin{equation}
\tilde J_3=J_3, \qquad p=\frac{J_-}{2\alpha}\,, \qquad 
q=\frac{\tilde J_+}{2\alpha}\,.
\label{75}\end{equation}
If we want an explicit single-valued map from $L(u)$ to $\tilde L(u)$
we must express $M(u,\lambda)$, and therefore $p$ and $q$, in term of the old variables. To
solve this problem
we use the spectrality of the BT. As well as the equation \Ref{71} that our map satisfies,
it will be parametrized
by a point $P=(\lambda,\mu) \in \Gamma$. Notice that there are two points on
$\Gamma$,
$P=(\lambda,\mu)$ and $Q=(\lambda,-\mu)$, corresponding to the same $\lambda$ and sitting
one above
the other because of the elliptic involution:
\begin{equation}
(\lambda,\mu) \in \Gamma: \qquad \det(L(\lambda) 
-\mu)=0 \qquad
\Leftrightarrow
\qquad \mu^2+ \det(L(\lambda))=0.
\end{equation}
As shown in \cite{HKR}, this spectrality property, used as a new datum, produces
the formula
\begin{equation}
q=\frac{A(\lambda)-\mu}{B(\lambda)}=-\frac{C(\lambda)}{A(\lambda)+\mu}\,,
\label{77}\end{equation}
where $\lambda$ and $\mu$ are bound by the equation for the elliptic curve
\begin{equation}
\mu^2=-\left( \frac{1}{\l^4}+\frac{2\ell}{\l^3}+\frac{2\mathcal{H}}{\l^2}+
\frac{2J_3}{\l}+\alpha^2 \right).
\end{equation}
Now the equation \Ref{71} gives an integrable Poisson map from $L(u)$ to $\tilde
L(u)$. The map
is parametrized by one point $(\lambda,\mu) \in \Gamma$. Explicitly, it reads
\begin{eqnarray}
\tilde J_3 &=& J_3, \label{79-84}\\
\tilde J_- &=& x_- +J_- (pq-\lambda)-2pJ_3, \nonumber\\
\tilde J_+ &=& 2q \alpha,\nonumber\\
\tilde x_3 &=& x_3 + pJ_+ -qx_- -qJ_- (pq- \lambda)+2pqJ_3, \nonumber\\
\tilde x_- &=& x_- (2pq- \lambda)-2px_3 -p^2 J_+
               +J_- pq(pq-\lambda)-2p^2 q J_3, \nonumber\\
\tilde x_+ &=& J_+ -J_- \frac{q}{p}(pq-\lambda)+2qJ_3.
\nonumber\end{eqnarray}
If we refer to the real generators of $e(3)$ the map takes the following explicit form,
\begin{eqnarray}
\tilde J_1 &=& \alpha q +\frac{1}{2}(x_1-ix_2) - \frac{\l}{2}(J_1-iJ_2)
+\frac{q}{4\alpha}
(J_1^2 - J_2^2)+ \label{85-90}\\
& & + \frac{i}{2 \alpha}(J_2J_3-q J_1J_2+iJ_1J_3), \nonumber\\
\tilde J_2 &=& -i \alpha q +\frac{i}{2}(x_1+ix_2) - \frac{i\l}{2}(J_1+iJ_2)
+\frac{iq}{4 \alpha}
(J_1^2-J_2^2)+ \nonumber \\
& & + \frac{1}{2 \alpha}(J_2J_3-q J_1J_2-iJ_1J_3), \nonumber\\
\tilde J_3 &=& J_3,
\nonumber\end{eqnarray}
\begin{eqnarray}
\tilde x_1 &=&\alpha q \l +\frac{1}{2}(J_1+iJ_2) - \frac{q^2}{2}(J_1-iJ_2) -
\frac{\l}{2}(x_1-ix_2) +qJ_3 + \nonumber \\
& & + \frac{q}{2 \alpha} \left[ \frac{\l}{2} (J_2^2-J_1^2)+(J_1-iJ_2) 
\left(x_1-\frac{x_3}{q} \right)
-ix_2 (J_1+iJ_2) + i \l J_1J_2 \right] + \nonumber \\
& & + \frac{q}{8 \alpha^2} \left[ q J_1^2 (J_1 -3iJ_2) + 
q J_2^2 (J_2 -3iJ_1) -
\frac{i}{q}(J_1-J_2)^2(J_1+J_2)-2J_3(J_1-iJ_2)^2 \right], \nonumber\\
\tilde x_2 &=& -i \alpha q \l +\frac{i}{2}(J_1-iJ_2) - 
\frac{iq^2}{2}(J_1+iJ_2) -
\frac{i\l}{2}(x_1-ix_2) -i qJ_3 + \nonumber \\
& & + \frac{i q}{2 \alpha} \left[ \frac{\l}{2} (J_2^2-J_1^2)+(J_1-iJ_2) 
\left(x_1-
\frac{x_3}{q} \right)
-ix_2 (J_1+iJ_2) + i \l J_1J_2 \right] + \nonumber \\
& & + \frac{i q}{8 \alpha^2} \left[ q J_1^2 (J_1 -3iJ_2) + q J_2^2 (J_2 -3iJ_1) -
\frac{i}{q}(J_1-J_2)^2(J_1+J_2)-2J_3(J_1-iJ_2)^2 \right], \nonumber\\
\tilde x_3 &=& x_3 -q(x_1-ix_2) +q \l (J_1 -iJ_2) + \nonumber \\
& & +  \frac{1}{\alpha} \left[ \frac{1}{2} (J_1^2 +J_2^2) - \frac{q^2}{2}
(J_1^2 -J_2^2) -iq
(J_2J_3-q J_1J_2+iJ_1J_3) \right] ,
\nonumber\end{eqnarray}
where we have used \Ref{75} to express $p$ in terms of the old variables.

Notice that the above one-point BT is a complex map, so it is a non-physical
B\"acklund transformation.
%%%%%%%%%%%%%%%%%%%%%%%%%%%%%%%%%%%%%%%%%%%%%%%%%%
\subsection{{Symplecticity of the one-point BT}} %%
%%%%%%%%%%%%%%%%%%%%%%%%%%%%%%%%%%%%%%%%%%%%%%%%%%
\noindent
We give a simple proof of symplecticity of the constructed map by finding an explicit
generating function of the corresponding canonical transformation from the
old to new variables.

First, because the Casimir functions do not change under the map,
\begin{equation}
x_{3}^{2}+x_+x_-=\tilde x_{3}^{2}+\tilde x_+\tilde x_-  =1,
\end{equation}
\begin{equation}
\frac{1}{2}(J_+x_-+J_-x_+)+J_3 x_3=
\frac{1}{2}(\tilde J_+\tilde x_- +\tilde J_-\tilde x_+)+
\tilde J_3 \tilde x_3 =\ell,
\end{equation}
we can exclude the variables $x_+,J_+$ and  $\tilde x_-,\tilde J_-$,
using the following substitutions:
\begin{equation}
x_+=\frac{1-x_{3}^{2}}{x_-}\,, \qquad \qquad J_+=\frac{2 \ell}{x_-}-
\frac{2 J_3x_3}{x_-}-
\frac{J_-}{x_{-}^{2}}\,(1-x_{3}^{2})\,,
\end{equation}
\begin{equation}
\tilde x_-=\frac{1-\tilde x_{3}^{2}}{\tilde x_+}\,, \qquad \qquad
\tilde J_-=\frac{2 \ell}{\tilde x_+}-\frac{2 \tilde J_3 \tilde x_3}
{\tilde x_+}-
\frac{\tilde J_+}{\tilde x_{+}^{2}}\,(1-\tilde x_{3}^{2}).
\end{equation}
Now we have only four (old and new) independent variables:
$x_-,x_3,J_-,J_3$ and  $\tilde x_+,\tilde x_3,\tilde J_+,\tilde J_3$.
We write the 1-point BT as a canonical transformation
defined by the generating function
$F_\lambda (x_-,J_- \mid \tilde x_+,\tilde J_+)$ written in terms of
$x_-,J_-$ and $\tilde x_+,\tilde J_+$:
\begin{eqnarray}
x_3 &=& ix_- \,\frac{\partial F_\lambda (x_-,J_- \mid \tilde x_+,\tilde J_+)}
{\partial J_-}\,,\label{95-98}\\
J_3 &=& ix_- \,\frac{\partial F_\lambda (x_-,J_- \mid \tilde x_+,\tilde J_+)}
{\partial x_-}+
iJ_- \,\frac{\partial F_\lambda (x_-,J_- \mid \tilde x_+,\tilde J_+)}
{\partial J_-}\,,\nonumber\\
\tilde x_3 &=&
i\tilde x_+ \,\frac{\partial F_\lambda (x_-,J_- \mid \tilde x_+,\tilde J_+)}
{\partial \tilde J_+}\,,\nonumber\\
\tilde J_3 &=&
i\tilde x_+ \,\frac{\partial F_\lambda (x_-,J_- \mid \tilde x_+,\tilde J_+)}
{\partial \tilde x_+}+
i\tilde J_+\, \frac{\partial F_\lambda (x_-,J_- \mid \tilde x_+,\tilde J_+)}
{\partial \tilde J_+}\,.
\nonumber\end{eqnarray}
With the help of \Ref{75} we rewrite
equations \Ref{79-84} of the map in the form
\begin{eqnarray}
x_3 &=& \frac{x_- \tilde J_+}{2\alpha} +k, \qquad\qquad k^2=1+\lambda x_-\tilde x_+,
\label{99-102}\\
J_3&=&\frac{\ell}{k}+\frac{\lambda}{2k}\,(J_- \tilde x_+ +
x_- \tilde J_+) -\frac{x_-\tilde x_+}{2k}+\frac{J_-\tilde J_+}{2\alpha}\,,\nonumber\\
\tilde x_3 &=& \frac{J_- \tilde x_+}{2\alpha} +k,\nonumber\\
\tilde J_3&=&J_3.
\nonumber\end{eqnarray}
It is now easy to check that the function
\begin{equation}
F_\lambda (x_-,J_- \mid \tilde x_+,\tilde J_+)=-i\,\frac{J_- \tilde J_+}{2\alpha}
-ik\left( \frac{J_-}{x_-}+
\frac{\tilde J_+}{\tilde x_+} -\frac{1}{\lambda} \right)
+i\ell \log{\frac{1+k}{1-k}}-i \alpha \l,
\label{104}\end{equation}
solves the equations \Ref{95-98}--\Ref{99-102}. Symplecticity of the map is therefore proven.

Alternatively, we can derive \Ref{104} directly from the 
generating function constructed for the $sl(2)$ Gaudin magnet 
in \cite{HKR}. The one-point BT in that case is given by
\begin{equation}
s_j^3 = i s_j^- \,
\frac{\partial F_\lambda (s_1^-,...,s_n^-\mid \tilde s_1{}^+,..., \tilde s_n{}^+)}
{\partial s_j^-}\,, \qquad 
\tilde s_j{}^3 = i \tilde s_j{}^+\, 
\frac{\partial F_\lambda (s_1^-,...,s_n^-\mid \tilde s_1{}^+,..., \tilde s_n{}^+)}
{\partial \tilde s_j{}^+}\,,
\end{equation}
where
\begin{equation}
F_\lambda (s_1^-,...,s_n^-\mid \tilde s_1{}^+,..., \tilde s_n{}^+)=
-\frac{i}{2\alpha} \sum_{j,k=1}^n s_j^- \tilde s_k{}^+
-i \sum_{j=1}^n \left(2z_j +s_j \log{\frac{z_j-s_j}{z_j +s_j }} \right) -i \alpha \lambda,
\label{Ggf}\end{equation}
\begin{equation}
z_j^2 = s_j^2 - (a_j -\lambda) \tilde s_j{}^+  s_j^-, \qquad 
s_j^2=(s_j^3)^2+s_j^- s_j^+ = (\tilde s_j{}^3)^2+\tilde s_j{}^- \tilde s_j{}^+, \qquad  
j=1,\ldots,n.
\end{equation}

Set $n=2$ in \Ref{Ggf} and impose the coalescence $a_2=a_1+\epsilon$, 
$\epsilon \rightarrow 0$, similar to the
derivation of the Lax matrix for the Lagrange top in Section 2. 
Now redefine $\lambda-a_1 \equiv \lambda$ and obtain
\begin{equation}
\mathcal{F}_\lambda (s_1^-,s_2^- \mid \tilde s_1{}^+,\tilde s_2{}^+)=
-\frac{i}{2\alpha} \sum_{j,k=1}^2 s_j^- \tilde s_k{}^+
-i \sum_{j=1}^2 \left(2 \zeta_j +s_j \log{\frac{\zeta_j-s_j}{\zeta_j +s_j }} \right) 
-i \frac{\zeta_2}{\lambda} \epsilon -i \alpha \lambda +O (\epsilon^2),
\label{conGF}\end{equation}
where
\begin{equation}
\zeta_j^2= s_j^2 + \lambda \tilde s_j{}^+  s_j^-, \qquad \qquad j=1,2.
\end{equation}
Referring to the formulae \Ref{29} we easily deduce that
\begin{equation}
\zeta_1^2 = s_1^2 + \lambda \left(J_- \tilde J_+ -\frac{\tilde J_+ x_- + J_- \tilde x_+}{\epsilon} -
\frac{x_- \tilde x_+}{\epsilon^2} \right), \quad 
s_1^2 = J_3^2 + J_+ J_-  -\frac{2 \ell}{\epsilon}+ \frac{1}{\epsilon^2} \,, 
\label{zs1}\end{equation}
\begin{equation}
\zeta_2^2 = s_2^2 (1+\lambda x_-\tilde x_+) , \qquad 
s_2^2 = \frac{1}{\epsilon^2}\,.
\label{zs2}\end{equation}
Now it is easy to check that \Ref{conGF} with \Ref{zs1}--\Ref{zs2} coincides 
with the generating function
\Ref{104}.

%%%%%%%%%%%%%%%%%%%%%%%%%%%%%%%%%%%%%%%%%%%%%%%%%%
\subsection{Spectrality of the one-point BT} %%
%%%%%%%%%%%%%%%%%%%%%%%%%%%%%%%%%%%%%%%%%%%%%%%%%
\noindent
The spectrality property of a B\"acklund transformation \cite{KS5}
means that the two components, $\lambda$ and $\mu$,
of the point $(\lambda,\mu) \in \Gamma$
parametrizing the map are conjugated variables, in
the sense that
\beq
\mu = -\frac{\partial F}{ \partial \l}\,,
\eeq
 where $F$ is the generating function of the BT.

We now show the spectrality property for the one-point BT constructed in the
previous section. Using the 
equations \Ref{75}, \Ref{77}, \Ref{99-102} and \Ref{104} we obtain
\begin{eqnarray}
\mu&=&A(\lambda)-\frac{\tilde J_+}{2\alpha}\,B(\lambda)= \frac{i}{k}
\left[  
\frac{1}{\l^2}+\frac{\ell}{\l} + \frac{x_- \tilde x_+}{2}\left(  \frac{J_-}{x_-}+
\frac{\tilde J_+}{\tilde x_+} -\frac{1}{\lambda} \right) + \alpha k
 \right] = \\
&= & -\frac{\partial F_\lambda (x_-,J_- \mid \tilde x_+,\tilde J_+)}
{\partial \lambda}\,.
\nonumber
\end{eqnarray}
%%%%%%%%%%%%%%%%%%%%%%%%%%%%%%%%%%%%%%%%%%%%%%%%%%
\subsection{{One-point BT in Eulerian variables}} %%
%%%%%%%%%%%%%%%%%%%%%%%%%%%%%%%%%%%%%%%%%%%%%%%%%%
\noindent
As shown in Section 4 one can represent the  Lie-Poisson 
algebra $e(3)$ using the canonical realization \Ref{59}. 
Now we rewrite the one-point BT as a 
canonical map in terms of these variables
\begin{equation}
\mathcal{B}_{\lambda}: \quad (\theta,\varphi, p_\theta,p_\varphi) \longmapsto
(\tilde \theta,\tilde \varphi, \tilde p_\theta, \tilde p_\varphi),
\end{equation}
where tilde  refers to the updated versions of the generators
\begin{eqnarray}
\tilde J_1&=&\tilde p_\theta\cos\tilde \varphi+\frac{\ell-\tilde p_\varphi\cos\tilde
\theta}
{\sin\tilde \theta}\,\sin\tilde \varphi,\qquad\quad
\tilde x_1=\sin\tilde \varphi\sin\tilde \theta,\\
\tilde J_2&=&-\tilde p_\theta\sin\tilde \varphi+\frac{\ell-\tilde p_\varphi\cos\tilde
\theta}
{\sin\tilde \theta}\,\cos\tilde \varphi,\qquad\,
\tilde x_2=\cos\tilde \varphi\sin\tilde \theta,\nonumber\\
\tilde J_3&=&\tilde p_\varphi,\qquad\qquad\qquad\qquad
\qquad\qquad\qquad \tilde x_3=\cos\tilde \theta,
\nonumber\end{eqnarray}
\begin{equation}
\{\tilde p_\theta,\tilde \theta\}=\{\tilde p_\varphi,\tilde \varphi\}=1.
\end{equation}
Recall that we have fixed the values of the two Casimir functions as follows:
$\sum_{k=1}^3x_k^2=1$ and $\sum_{k=1}^3 x_k J_k =\ell$. 
Using  the equations
\Ref{99-102} we can write down the one-point BT as:
\begin{eqnarray}
\qquad\;\; p_\varphi &=& \frac{1}{k} \left[ \ell +\lambda \alpha (\cos \theta + \cos \tilde
\theta )
+ \frac{k^2-1}{2 \lambda} \right] + 2 \alpha \lambda \left[ \frac{(\cos \theta-1)
(\cos \tilde \theta-1)}{k^2-1} -1 \right],  \label{111}\\
\tilde p_\varphi &=&p_\varphi, \nonumber\\
p_\theta &=& \frac{i}{\sin \theta}\,(\ell -  p_\varphi\cos \theta)
-\frac{2i \alpha \lambda \sin \theta (\cos \tilde \theta-k)}{k^2-1}\,,\label{113}\\
\tilde p_\theta &=& \frac{i}{\sin \tilde \theta}\,(p_\varphi\cos \tilde \theta - \ell)+
\frac{2i \alpha \lambda \sin \tilde \theta (\cos \theta-k)}{k^2-1}\,,
\nonumber\end{eqnarray}
with
\begin{equation}
k^2=1+\lambda \phi \tilde \phi \sin \theta \sin \tilde \theta, \qquad \phi \equiv
e^{i \varphi}, ~ \;\;
\tilde \phi \equiv e^{-i \tilde \varphi}.
\label{115}\end{equation}
Remark that in the last formula we have introduced two new variables $\phi$ and $\tilde \phi$: it is a
convenient choice
since the map depends only on
$e^{i\varphi}$ and
$e^{-i \tilde \varphi}$.

We want to focus our attention upon the symmetry of the canonical transform
\Ref{111}--\Ref{113},
which is symmetric under the exchange $(\theta, \varphi)
\leftrightarrow (-\tilde \theta, -\tilde \varphi)$. Notice that
the $\phi, \tilde \phi$ dependence is trivial: it involves only the product $\phi
\tilde \phi$
(i.e. $\varphi- \tilde \varphi$),
which is symmetric under the change $\phi \leftrightarrow \tilde \phi$.
The dependence upon $\theta, \tilde \theta$ is more interesting. Let us introduce
the exchange operator $\mathcal{P}_{\theta, \tilde \theta}$ such that
\begin{equation}
\mathcal{P}_{\theta, \tilde \theta} f(\theta, \tilde \theta)=f(\tilde \theta,\theta),
\qquad
\mathcal{P}_{\theta, \tilde \theta}^2=1,
\end{equation}
where $f(\theta, \tilde \theta)$ is a generic function. For instance we notice that
$\mathcal{P}_{\theta, \tilde \theta} k=k$, where $k$ is the function in \Ref{115}.
If we consider formulae \Ref{111}--\Ref{113} it is easy to notice that
\begin{equation}
\mathcal{P}_{\theta, \tilde \theta}\,p_\varphi=p_\varphi, \qquad
\mathcal{P}_{\theta, \tilde \theta}\,p_\theta=-\tilde p_\theta.
\label{117}\end{equation}
Our aim is to write down the canonical transform in the following form:
\begin{equation}
p_\theta=\frac{\partial F_\l(\theta,\phi|\tilde \theta,\tilde
\phi)}{\partial\theta}\,,
\quad
\tilde p_\theta=-\frac{\partial F_\l(\theta,\phi|\tilde \theta,\tilde
\phi)}{\partial\tilde\theta}\,,
\end{equation}
\be
p_\varphi=
i \phi \frac{\partial F_\l(\theta,\phi|\tilde \theta,\tilde \phi)}{\partial\phi}\,,
\quad \tilde p_\varphi=i \tilde \phi \frac{\partial F_\l(\theta,\phi|\tilde
\theta,\tilde \phi)}
{\partial\tilde \phi}\,,
\ee
where $F_\l(\theta,\phi|\tilde \theta,\tilde \phi)$ is the generating function of the
one-point BT. Since equations \Ref{117} hold the immediate consequence is that
\begin{equation}
F_\l(\theta,\phi|\tilde \theta,\tilde \phi)=\mathcal{P}_{\theta, \tilde \theta}
F_\l(\theta,\phi|\tilde \theta,\tilde \phi)=
F_\l(\tilde \theta,\tilde \phi|\theta,\phi).
\end{equation}
It is possible to prove that the generating function is:
\begin{equation}
F_\l(\theta,\phi|\tilde \theta,\tilde \phi)= -i \alpha \l (2\ln \phi \tilde \phi +1)
+A_\l(\theta,\phi|\tilde \theta,\tilde \phi)+
\mathcal{P}_{\theta, \tilde \theta}A_\l(\theta,\phi|\tilde \theta,\tilde \phi),
\end{equation}
where
\begin{equation}
A_\l(\theta,\phi|\tilde \theta,\tilde \phi) =
a_0+a_1 E(\xi | \eta)+a_2 F(\xi | \eta)+a_3 \Pi(\zeta; \xi | \eta)
\end{equation}
with
\begin{equation}
\left\{ \begin{array}{llll}
a_0&=& i \ell \ln \left(\frac{\sin \theta}{\cos \theta +1}\right)-i\frac{k}{2 \l}
+\frac{1}{2} i \ell \ln \left(\frac{1+k}{1-k}\right)
+2 i \alpha \l \left\{ \ln(\sin \theta)(1+ \cos \tilde \theta)-k +  \right.\\
& & \left. +\frac{k^2-2}{k^2-1} \cos \theta+ \frac{k}{2(k^2-1)}(1+\cos \theta \cos
\tilde \theta)+
\frac{\theta \sin \theta \sin^2 \tilde \theta}{(k^2-1)(\cos \tilde \theta -1)}
+ \ln \left(\frac{\sin \theta}{\cos \theta +1}\right)+ \right. \\
& & + \left.\frac{1}{2} \cos \theta \ln \left(\frac{1+k}{1-k}\right)+ \frac{3}{4}
\ln \left(\frac{k+1}{k-1}\right)\right\},\\
& & \\
a_{1}&=& \frac{2i \alpha S}{k\cos \theta} \left[(1+\cos \tilde \theta) \left(3 \l -
\frac{1}{\phi^2 \tilde \phi^2 \sin^2 \tilde \theta}\right) -
\frac{2}{\l \phi^2 \tilde \phi^2 (1- \cos \tilde \theta)}\right],\\
& & \\
a_{2}&=& \frac{2i \alpha S}{k\cos \theta} \left[
\frac{2}{\l \phi^2 \tilde \phi^2 (1- \cos \tilde \theta)}-
(1+\cos \tilde \theta) \left(3 \l +
\frac{1}{\phi \tilde \phi \sin\tilde \theta}\right) \right],\\
& & \\
a_{3}&=& 2 i \alpha \l S \frac{(1- \cos \tilde \theta)(k^2-1+\sin \theta)}{k\cos
\theta \sin \theta}\,,\\
\end{array} \right.
\end{equation}
\begin{equation}
\left\{ \begin{array}{lll}
E(\xi | \eta)&=& \int_0^\xi (1- \sin^2 \eta \sin^2 z)^{1/2} dz,\\
& & \\
F(\xi | \eta)&=& \int_0^\xi (1- \sin^2 \eta \sin^2 z)^{-1/2} dz,\\
& & \\
\Pi(\zeta; \xi | \eta)&=& \int_0^\xi (1- \zeta^2 \sin^2 z)^{-1}
(1- \sin^2 \eta \sin^2 z)^{-1/2} dz,\\
\end{array} \right.
\end{equation}
\begin{equation}
\left\{ \begin{array}{llll}
S&=&\frac{k \cos \theta (k^2-1)}{k^2-2k + \cos^2 \theta} \sqrt{\sin \theta (k-1-\sin\theta)}\,,\\
& & \\
\xi &=& k \sqrt{\frac{\sin \theta}{k^2-1+\sin \theta}}\,,\\
& & \\
\eta &=& \sqrt{-\frac{k^2-1+\sin \theta}{k^2-1-\sin \theta}}\,,\\
& & \\
\zeta &=&\frac{k^2-1+\sin \theta}{\sin \theta}\,.
\end{array} \right.
\end{equation}
In the zero-field case ($\alpha=0$) the generating function simply reads:
\begin{equation}
F_\l^{(\alpha=0)}(\theta,\phi|\tilde \theta,\tilde \phi)=i \ell \ln
\left[ 
\frac{1+k}{1-k}\frac{\sin \theta \sin \tilde \theta}{(\cos \theta+1)(\cos \tilde \theta+1)}
\right] -i \,\frac{k}{\l}\,.
\end{equation}
%%%%%%%%%%%%%%%%%%%%%%%%%%%%%%%%%%%%%%%%%%%%%%%%%%
\subsection{Two-point BT}
\label{5.5}
%%%%%%%%%%%%%%%%%%%%%%%%%%%%%%%%%%%%%%%%%%%%%%%%%%
\noindent
According to \cite{HKR}, we now construct a composite map which is a product of the map
$\mathcal{B}_{P_1} \equiv \mathcal{B}_{(\lambda_1,\mu_1)}$ and
$\mathcal{B}_{Q_2} \equiv \mathcal{B}_{(\lambda_2,-\mu_2)}$:
\begin{equation}
\mathcal{B}_{P_1,Q_2}=\mathcal{B}_{Q_2} \circ \mathcal{B}_{P_1}:~
L(u) \stackrel{\mathcal{B}_{P_1}} {\longmapsto} \tilde L(u)
\stackrel{\mathcal{B}_{Q_2}} {\longmapsto} \stackrel{\approx}{L}(u).
\end{equation}
Two maps are inverse to each other when $\lambda_1=\lambda_2$ and $\mu_1=\mu_2$.
This  two-point BT 
for the Lagrange top is defined by the following `discrete-time' Lax equation:
\begin{equation}
M(u;\lambda_1,\lambda_2)L(u)=\stackrel{\approx}{L}(u)M(u;\lambda_1,\lambda_2) \qquad \forall u, \qquad \lambda_1,\lambda_2 \in \mathbb{C},
\label{128}\end{equation}
where the matrix $M(u;\lambda_1,\lambda_2)$ is
\begin{equation}
M(u;\lambda_1,\lambda_2)=\begin{pmatrix}u-\l_1+xX&X\cr -x^2X+(\l_1-\l_2)x&u-\l_2-xX\end{pmatrix},
\label{129}\end{equation}
\begin{equation}
\det M(u;\lambda_1,\lambda_2) =(u - \l_1)(u - \l_2).
\end{equation}
Spectrality property with respect to two fixed points $(\lambda_1,\mu_1)\in \Gamma$
and $(\lambda_2,\mu_2)\in \Gamma$ give
\begin{equation}
x=\frac{A(\lambda_1)-\mu_1}{B(\lambda_1)}=- \frac{C(\l_1}{A(\l_1)+ \mu_1}
=\frac{\stackrel{\approx}{A}(\lambda_2)-\mu_2}
{\stackrel{\approx}{B}(\lambda_2)}=-\frac{\stackrel{\approx}{C}(\lambda_2)}
{\stackrel{\approx}{A}(\lambda_2)
+\mu_2}
=\frac{\tilde J_+}{2 \alpha}\,,
\label{ex1}\end{equation}
\begin{eqnarray}
X&=&\frac{(\lambda_2-\lambda_1)B(\lambda_1)B(\lambda_2)}
{B(\lambda_1)(A(\lambda_2)+\mu_2)-B(\lambda_2)(A(\lambda_1)-\mu_1)}
\label{ex2}\\
&=&\frac{(\lambda_1-\lambda_2)(A(\lambda_1)+\mu_1)(A(\lambda_2)-\mu_2)}
{(A(\lambda_1)+\mu_1)C(\l_2)-(A(\lambda_2)-\mu_2)C(\l_1)}\nonumber\\
&=&\frac{(\lambda_2-\lambda_1)\stackrel{\approx}{B}(\lambda_1)
\stackrel{\approx}{B}(\lambda_2)}
{\stackrel{\approx}{B}(\lambda_2)(\stackrel{\approx}{A}(\lambda_1)+\mu_1)-
\stackrel{\approx}{B}(\lambda_1)(\stackrel{\approx}{A}(\lambda_2)-\mu_2)} \nonumber\\
&=&\frac{(\lambda_1-\lambda_2)(\stackrel{\approx}{A}(\lambda_1)-\mu_1)
(\stackrel{\approx}{A}(\lambda_2)+\mu_2)}
{(\stackrel{\approx}{A}(\lambda_2)+\mu_2)\stackrel{\approx}{C}(\l_1)-
(\stackrel{\approx}{A}(\lambda_1)-\mu_1)\stackrel{\approx}{C}(\l_2)} \nonumber\\
&=& \frac{J_- - \stackrel{\approx}{J}_-}{2 \alpha}\,.
\nonumber\end{eqnarray}
Now we have two 
 B\"acklund parameters $\lambda_1,\lambda_2 \in \mathbb{C}$. The above formulae
give several equivalent
expressions for the variables $x$ and $X$ since the points
$(\lambda_1,\mu_1)$ and $(\lambda_2,\mu_2)$ belong to the spectral curve $\Gamma$,
i.e., are bound
by the following relations
\begin{equation}
\mu_k^2=A^2(\lambda_k)+B(\lambda_k)C(\lambda_k)=
\stackrel{\approx}{A}^2(\lambda_k)+\stackrel{\approx}{B}(\lambda_k)\stackrel{\approx}{C}(\lambda_k),
\qquad k=1,2.
\end{equation}

Together with \Ref{ex1}--\Ref{ex2}, the 
 formula \Ref{128} give an explicit two-point Poisson integrable map from $L(u)$ to
$\stackrel{\approx}{L}(u)$ (as well as its inverse, i.e the map from 
$\stackrel{\approx}{L}(u)$ to $L(u)$).
The map is parametrized by two points $\mathcal{B}_{P_1} \equiv
\mathcal{B}_{(\lambda_1,\mu_1)}$ and
$\mathcal{B}_{Q_2} \equiv \mathcal{B}_{(\lambda_2,-\mu_2)}$.
Let us introduce the notation
\begin{equation}
\Delta \l = \l_1 - \l_2, \qquad \qquad \l_0 = \frac{\l_1+ \l_2}{2}\,.
\end{equation}
The map explicitly reads:
\beq
\stackrel{\approx}{J}_3 =J_3,\qquad
\stackrel{\approx}{J}_- = J_- -2X,\qquad
\stackrel{\approx}{J}_+ = J_+
+2x (\Delta \lambda - xX), 
\eeq
\begin{eqnarray}
\stackrel{\approx}{x}_3 &=& x_3 -
\alpha (\Delta \lambda - xX)J_- +XJ_+ +
2xX(\Delta \lambda - xX),\nonumber\\
\stackrel{\approx}{x}_- &=& x_- +(2xX+\Delta \lambda)J_- -2X J_3    -
2X\left(xX +\l_0 - \frac{\Delta \l}{2} \right),\nonumber\\
\stackrel{\approx}{x}_+ &=& x_+ -(2xX+\Delta \lambda)J_+
+2x (\Delta \lambda - xX)J_3
-2x(\Delta \lambda - xX)\left(xX - \l_0 - \frac{\Delta \l}{2} \right).
\nonumber
\end{eqnarray}
It is useful to write down the B\"acklund map in terms of  the real generators of
$e(3)$:
\beq
\stackrel{\approx}{J}_1=J_1+x \Delta \l -X(1+x^2),\qquad
\stackrel{\approx}{J}_2=J_2-i \left[x \Delta \l + X (1-x^2)\right],\qquad
\stackrel{\approx}{J}_3=J_3,
\label{145-150}\eeq
\begin{eqnarray}
\stackrel{\approx}{x}_1&=&
x_1-iJ_2 (2xX-\Delta \l)
+J_3 \left[x \Delta \l -X(1+x^2)\right]- \nonumber \\
& & -\left\{ xX^2(1+x^2) +X \left[ \l_0 -\frac{\Delta \l}{2}-x^2\left(
\l_0+ \frac{3 \Delta \l}{2} \right) \right]  +x \Delta \l \left(\l_0 + \frac{\Delta \l}{2}
 \right)\right\},\nonumber\\
\stackrel{\approx}{x}_2&=&
x_2+iJ_1(2xX-\Delta \l)-iJ_3 \left[ x \Delta \l +X (1-x^2) \right] - \nonumber \\
& & -i \left\{ xX^2(1-x^2) +X \left[ \l_0 -\frac{\Delta \l}{2}+x^2 \left( 
\l_0+ \frac{3 \Delta \l}{2} \right) \right] -x \Delta \l \left(\l_0 + \frac{\Delta \l}{2}
 \right) \right\},\nonumber\\
\stackrel{\approx}{x}_3&=&x_3-J_1 \left[x \Delta \l -X(1+x^2) \right]+iJ_2 
\left[x \Delta \l +X (1-x^2) \right]  -2 xX(xX - \Delta \l).
\nonumber\end{eqnarray}
Obviously, when $\lambda_1=\lambda_2$ (and $\mu_1=\mu_2$) the map turns
into an identity map. 
With a direct calculation one can show that the map \Ref{145-150} sends real variables
to real variables provided
\begin{equation}
\lambda_1=\bar\lambda_2 \equiv \lambda \in \mathbb{C}.
\end{equation}
Therefore,  the two-point map leads to a physical B\"acklund transformation
with two real parameters.

%%%%%%%%%%%%%%%%%%%%%%%%%%%%%%%%%%%%%%%%%%%%%%%%%%
\subsection{Two-point BT as a discrete-time map} %%
%%%%%%%%%%%%%%%%%%%%%%%%%%%%%%%%%%%%%%%%%%%%%%%%%%

\noindent
The two point BT constructed above is a one-parameter
($\lambda_1$) time discretization
 of a family of flows parametrized by the point $Q_2=(\lambda_2,-\mu_2)$, with the
difference
$\Delta \l$ playing the role of a time step. Recall that the physical time step is
$i\Delta \l$ since
$\Delta \l \in i \mathbb{R}$.

Consider the limit
\begin{equation}
i \Delta \l = i(\lambda_1-\lambda_2)= \epsilon, \qquad \epsilon \rightarrow 0.
\end{equation}
It is easy to see from the formulae of the previous section that
\begin{eqnarray}
x&=& x_0 +O(\epsilon), \qquad
x_0=\frac{A(\lambda_2)-\mu_2}{B(\lambda_2)}= -\frac{C(\l_2)}{A(\lambda_2)+\mu_2}\,, \\
X &=& \epsilon X_0 +O(\epsilon^2), \qquad
X_0 =\frac{iB(\lambda_2)}{2\mu_2}\,.
\nonumber\end{eqnarray}
The matrix $M$ has the following asymptotics:
\begin{equation}
M(u;\lambda_2)=(u-\lambda_2)\left[1
+\frac{i \epsilon}{2\mu_2(u-\lambda_2)}
\left(\begin{array}{cc}
A(\lambda_2)+\mu_2 & B(\lambda_2) \\
C(\lambda_2) & -A(\lambda_2)+\mu_2
\end{array}\right)
\right]
+O(\epsilon^2).
\end{equation}
If we define the time derivative $\dot L(u) $ as
\begin{equation}
\dot L(u) \equiv \lim_{\epsilon \rightarrow 0}
\frac{\stackrel{\approx}{L}(u)-L(u)}{\epsilon}\,,
\end{equation}
then in this limit we obtain from the equation
$M(u)L(u)=\stackrel{\approx}{L}(u)M(u)$ the Lax
equation for the  corresponding continuous flow that our BT discretizes, namely:
\begin{equation}
\dot L(u) =\frac{i}{2 \mu_2} \left[
\frac{L(\lambda_2)}{(u-\lambda_2)}, L(u)
\right].
\label{160}\end{equation}
This is a Hamiltonian flow with $i\mu_2$,
\begin{equation}
i\mu_2 =\sqrt{-A(\lambda_2)^2-B(\lambda_2)C(\lambda_2)} =\sqrt{
\frac{1}{\lambda_2^4}+\frac{2\ell}{\lambda_2^3}+
\frac{2\mathcal{H}}{\lambda_2^2}+\frac{2J_3}{\lambda_2}+\alpha^2},
\end{equation}
as the Hamiltonian function,
\begin{equation}
\dot L_{ij}(u) = i\{\mu_2, L_{ij}\}.
\end{equation}

Hence, the constructed two-point BT discretizes a one-parameter family of flows,
labelled by the arbitrary value of $\lambda_2$. Now,
let us consider
the following limit:
\begin{equation}
\lambda_2 = i \eta, \qquad \eta \rightarrow 0.
\end{equation}
In this limit we obtain the following expression for $\mu_2$:
\begin{equation}
\mu_2=\frac{i}{\eta^2}-\frac{\ell}{\eta}+i \left(\frac{\ell^2}{2}-\mathcal{H}
\right) +O(\eta).
\end{equation}
The Lax equation \Ref{160} turns into:
\begin{equation}
\dot L(u)=\frac{i}{2 u} \left[ \left(\begin{array}{cc}
x_{3} & x_{-}  \\
x_{+} & -x_{3}
\end{array} \right), L(u) \right]= \left(\begin{array}{cc}
\frac{J_- x_+ - J_+ x_-}{2u^2} & \frac{J_3 x_- - J_- x_3}{u^2}+ \frac{\alpha
x_-}{u}\\
\frac{J_+ x_3 - J_3 x_+}{u^2}- \frac{\alpha x_+}{u} & \frac{J_+ x_- - J_- x_+}{2u^2}
\end{array} \right),
\end{equation}
which is the flow with the Hamiltonian $\mathcal{H}$ \Ref{H}: $\dot L_{ij}(u)=\{\mathcal{H},L_{ij}(u)\}$.

It is possible to show that one iteration of the constructed two-point BT
is equivalent to moving from time 0 to time 1 along the flow with
the Hamiltonian (interpolating flow):
\beq
H(\lambda_1,\lambda_2)= i \int_{\lambda_1}^{\lambda_2} \mu(\lambda) d \lambda.
\eeq

%%%%%%%%%%%%%%%%%%%%%%%%%%%%%%%%%%%%%%%%%%%%%%%%%%
\subsection{Numerics}
\setcounter{equation}{0} %%
%%%%%%%%%%%%%%%%%%%%%%%%%%%%%%%%%%%%%%%%%%%%%%%%%%
\noindent Below is the MATLAB program, which shows
the real 2-point map from the Section \ref{5.5}.

{\tiny
\begin{verbatim}
function bt3 = l(aa,lambdar,lambdai,N,JJ1,JJ2,JJ3,xx1,xx2,xx3)
alpha=double(aa); 
lambda1=double(complex(double(lambdar),double(lambdai)));
lambda2=double(conj(lambda1));
J3(1)=double(JJ3); 
Jp(1)=double(complex(double(JJ1),double(JJ2))); 
Jm(1)=double(conj(Jp(1)));
x3(1)=double(xx3); 
xp(1)=double(complex(double(xx1),double(xx2))); 
xm(1)=double(conj(xp(1)));
C1=double(x3(1)^2+xp(1)*xm(1)); 
C2=double(xp(1)*Jm(1)+xm(1)*Jp(1)+2*x3(1)*J3(1));
H =double(J3(1)^2+Jp(1)*Jm(1)+2*alpha*x3(1));
mu1=double(sqrt(double(alpha^2+2*alpha*J3(1)/lambda1+H/lambda1^2+C2/lambda1^3+C1/lambda1^4)));
mu2=double(conj(mu1));
L0=[alpha 0; 0 -alpha];
LJ(1,1,1)=J3(1); LJ(1,2,1)=Jm(1); LJ(2,1,1)=Jp(1); LJ(2,2,1)=-J3(1); 
Lx(1,1,1)=x3(1); Lx(1,2,1)=xm(1); Lx(2,1,1)=xp(1); Lx(2,2,1)=-x3(1); 
for k=2:N
   x=double(((alpha-mu1)*lambda1^2+J3(k-1)*lambda1+x3(k-1))/(Jm(k-1)*lambda1+xm(k-1)));
   dd1=double((Jm(k-1)*lambda1+xm(k-1))*((alpha+mu2)*lambda2^2+J3(k-1)*lambda2+x3(k-1)));
   dd2=double((Jm(k-1)*lambda2+xm(k-1))*((alpha-mu1)*lambda1^2+J3(k-1)*lambda1+x3(k-1)));   
   X=double(double((lambda2-lambda1)*(Jm(k-1)*lambda1+xm(k-1))*(Jm(k-1)*lambda2+xm(k-1)))/double(dd1-dd2));
   M0=[double(-lambda1+x*X) X; 
           double(-x^2*X+(lambda1-lambda2)*x) double(-lambda2-x*X)];
   LJ(:,:,k)=double(LJ(:,:,k-1)+M0*L0-L0*M0);
   Lx(:,:,k)=double(Lx(:,:,k-1)+M0*LJ(:,:,k-1)-LJ(:,:,k)*M0);
   J3(k)=LJ(1,1,k); x3(k)=Lx(1,1,k); Jm(k)=LJ(1,2,k); xm(k)=Lx(1,2,k); 
end
bt3=plot3(real(xm),real(i*xm),real(x3),':oy','LineWidth',0.5,'MarkerEdgeColor',
'b','MarkerSize',1.5,'MarkerFaceColor','b');
axis('equal')
\end{verbatim}}

The input parameters are:
\begin{itemize}
\item aa$\,=\,\alpha$,
\item lambdar$\,=\,\Re \lambda$,
\item lambdai$\,=\,\Im \lambda$,
\item N\,=\,number of iterations of the map,
\item
JJ1,JJ2,JJ3,xx1,xx2,xx3\,=\,initial values of $J_1,J_2,J_3,x_1,x_2,x_3$.
\end{itemize}
The output is a 3D plot of $N$ consequent (blue) points $(x_1,x_2,x_3)$ linked
by the yellow lines. Notice that all these points lie on the sphere $x_1^2+x_2^2+x_3^2=$\,constant,
of some radius defined by the initial data.

\begin{figure}[h!]

\begin{center}
\includegraphics[height=10cm]{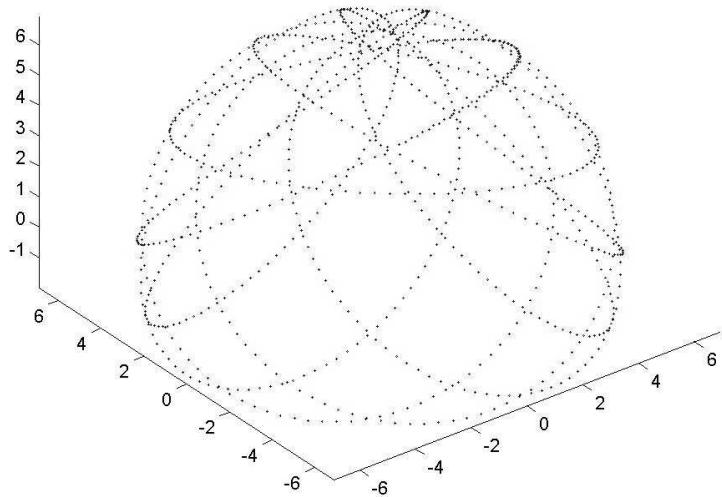}
\end{center}
\caption{$\scriptstyle (\alpha;\Re \lambda,\Im \lambda;N;J_1,J_2,J_3,x_1,x_2,x_3)=
(3.345;0,0.02;1000;2.34,6.4,8,1.219,-0.78,6.77)$}

\end{figure}

\begin{figure}[h!]

\begin{center}
\includegraphics[height=10cm]{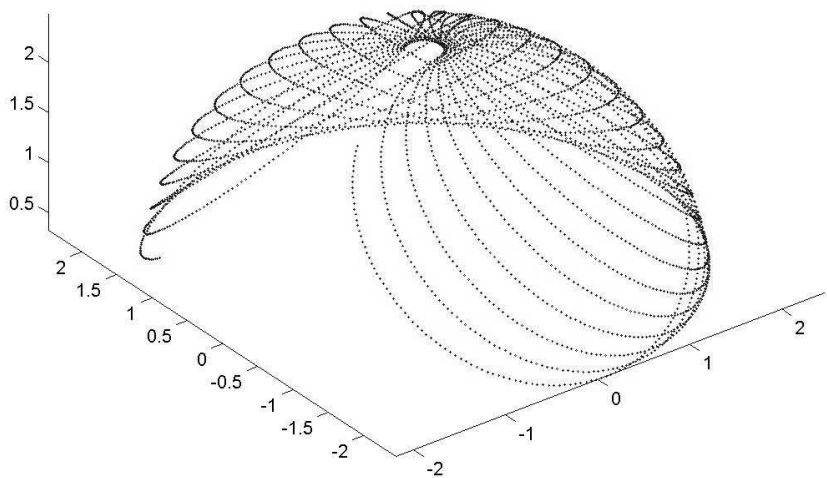}
\end{center}
\caption{$\scriptstyle(\alpha;\Re \lambda,\Im \lambda;N;J_1,J_2,J_3,x_1,x_2,x_3)=
(1;500,5000;4000;3.4,-4.6,-6.2,-2.19,0.89,0.77)$}

\end{figure}

%%%%%%%%%%%%%%%%%%%%%%%%%%%%%%%%%%%%%%%%%%%%%%%%%
\section{Concluding remarks}
%%%%%%%%%%%%%%%%%%%%%%%%%%%%%%%%%%%%%%%%%%%%%%%%

We have constructed the 1- and 2-point B\"acklund transformations for the symmetric Lagrange
top following the approach of \cite{KS5}. The application of the constructed maps as exact
numerical integrators of the continuous flows is considered.
As shown in Section 2 the Lagrange top has the same {\it linear} algebraic Poisson structure that
belongs to the $sl(2)$ Gaudin magnet. This allowed us to choose the same ansatz for the matrix $M(u)$
\Ref{ansatz} that has been used in \cite{HKR}.

Notice that we had the following algebraic situation, as in Gaudin models:
a Lax matrix \Ref{lax-matrix2} which satisfies a {\it linear} $r$-matrix algebra \Ref{14}
requires a $M$ matrix which comes from a {\it quadratic} $r$-matrix algebra \Ref{74}. 
Recall that in the Toda lattice 
\cite{KS5} and in the DST model \cite{KSS98} both
$L$ and $M$ are derived from the same quadratic $r$-matrix algebra \Ref{74}. This fact
shall show up in the construction of the quantum analogue of these B\"acklund
transformations.
Indeed an interesting problem is the quantization of the constructed maps, namely the problem
of Baxter $\mathbb{Q}$-operators for the symmetric Lagrange top. Actually this problem leads
to the study of product formulae for Heun functions. 
This work is in progress and the results will be reported in a separate paper.

%%%%%%%%%%%%%%%%%%%%%%%%%%%%%%%%%%%%%%%%%%%%%%%%%
\section*{Acknowledgments}
%%%%%%%%%%%%%%%%%%%%%%%%%%%%%%%%%%%%%%%%%%%%%%%%

The author MP wishes to acknowledge the support from the INFN (Istituto Nazionale di Fisica Nucleare).
The hospitality of the School of Mathematics, University of Leeds, extended to MP during his
visit to Leeds in 2003, where part of this work was done, is also kindly acknowledged.

%%%%%%%%%%%%%%%%%%%%%%%%%%%%%%%%%%%%%%%%%%%%%%%%%
%% Bibliography
%%%%%%%%%%%%%%%%%%%%%%%%%%%%%%%%%%%%%%%%%%%%%%%%


\begin{thebibliography}{40}
\bibitem {Ar} V.I. Arnold, {\it{Mathematical methods of classical mechanics}}, 2nd edition,
Springer (1989).
\bibitem {Au} M. Audin, {\it{Spinning tops}}, Cambridge University Press (1996).
\bibitem {Gav1} L. Gavrilov, {\it{ Generalized Jacobians of spectral curves
and completely integrable systems}}, Math. Z. {\bf 230} (1999), no. 3, 487--508.
\bibitem {Gav2} L. Gavrilov, A. Zhivkov, {\it{The complex geometry
of the Lagrange top}}, Enseign. Math. (2) {\bf 44} (1998), no. 1-2, 133--170.
\bibitem {G1} M. Gaudin, {\it{Diagonalisation d'une classe d' hamiltoniens de spin}},
J. de Physique {\bf 37} (1976) 1087--1098.
\bibitem {SB} A.I. Bobenko and Yu. B. Suris, {\it{Discrete time Lagrangian mechanics on
Lie groups, with an application to the Lagrange top}}, 
Commun. Math. Phys. {\bf 204} (1999) 147--188.
\bibitem {G2} M. Gaudin, {\it{La fonction d' onde de Bethe}}, Masson, Parigi (1983).
\bibitem{HKR} A.N.W.~Hone, V.B.~Kuznetsov and O.~Ragnisco,
{\it B\"{a}cklund transformations for the $sl_2$ Gaudin magnet},
Journal of  Physics A {\bf 34} (2001), 2477--2490.
\bibitem {IW} E. In\"on\"u and E.P. Wigner, {\it{On the contraction of groups ant their representations}},
Proc. Natl Acad. Sci. USA {\bf 39} 510--24.
\bibitem{K} V.B. Kuznetsov, {\it Separation of variables for the $\mathcal {D}_n$ type periodic Toda 
lattice}, J. Phys. A: Math. Gen. {\bf 30} (1997), 2127--2138.
\bibitem{KNS} V.B. Kuznetsov, F.W. Nijhoff and E.K. Sklyanin, {\it 
Separation of variables for the  Ruijsenaars system}, 
Commun. Math. Phys. {\bf 189} (1997), 855--877.
\bibitem{KSS98} V.B.~Kuznetsov, M.~Salerno and E.K.~Sklyanin,
{\it Quantum \bt\ for DST dimer model,} J.\ Phys.\ A: Math.\ Gen.
{\bf 33} (2000) 171--189
\bibitem{KS5} V.B.~Kuznetsov and E.K.~Sklyanin, {\it On \bt{s} for many-body
systems,} J.\ Phys.\ A: Math.\ Gen.\ {\bf 31} (1998) 2241--2251.
\bibitem{KV00} V.B.~Kuznetsov and P.~Vanhaecke,
{\it B\"acklund transformations for finite-di\-men\-si\-o\-nal integrable
systems: a geometric approach}, Journal of Geometry and Physics {\bf 44} (2002), 1--40.
\bibitem{Skl38} E.K.~Sklyanin, {\it Separation of variables. New trends}
Progr.\ Theor.\ Phys.\ Suppl.\ {\bf 118} (1995) 35--60.
\bibitem {SU} Yu. B. Suris, {\it{The problem of integrable discretization: hamiltonian
approach}}, Birkh\"auser Verlag (2003).
\end{thebibliography}
\end{document}